\documentclass{ragtime} 
\usepackage{graphicx}
\title[Magnetic null points in the magnetosphere of a plunging neutron star]
      {Emergence of magnetic null points in electro-vacuum magnetospheres of compact objects: The case of a plunging neutron star}

\author[O. Kop\'{a}\v{c}ek,
        T. Tahamtan    
        and V. Karas]
       {Ond\v{r}ej Kop\'{a}\v{c}ek\at{1,a} 
        Tayebeh Tahamtan\at[]{2} 
        and Vladim\'{i}r Karas\at[]{1}\\
        \ins{1}Astronomical Institute, Czech Academy of Sciences,\splitins[1]
        Bo\v{c}n\'{\i} II 1401, Prague, CZ-141 31, Czech Republic\\
        \ins{2}Institute of Theoretical Physics, Faculty of Mathematics and Physics, Charles University,\splitins[1] V~Hole\v{s}ovi\v{c}k\'ach 2, Prague, CZ-180~00, Czech Republic\\ \ins{a}\Email{kopacek@ig.cas.cz}}

\providecommand{\dif}{\mathrm{d}} 

\begin{document}
\begin{abstract}
Relativistic effects of compact objects onto electromagnetic fields in their vicinity are investigated using the test-field approximation. In particular, we study the possible emergence of magnetic null points which are astrophysically relevant for the processes of magnetic reconnection. While the magnetic reconnection occurs in the presence of plasma and may lead to violent mass ejection, we show here that strong gravitation of the supermassive black hole may actively support the process by suitably entangling the field lines even in the electro-vacuum description. In this  contribution we further discuss the case of a dipole-type magnetic field of the neutron star on the plunging trajectory to the supermassive black hole. While we have previously shown that given model in principle admits the formation of magnetic null points, here we explore whether and where the null points appear for the astrophysically relevant values of the parameters.
\end{abstract}

\section{Introduction}\label{intro}
Strong gravity may significantly influence the structure of the electromagnetic fields. On the other hand, the electromagnetic field contributes to the stress energy tensor $T_{\mu\nu}$, which constitutes the source term in the Einstein field equations, and thus affects the geometry of the spacetime. In general, we need to solve coupled Einstein-Maxwell equations to determine the geometry given by the metric $g_{\mu\nu}$ and the electromagnetic field described by the tensor $F_{\mu\nu}$. Nevertheless, the field intensities encountered in the astrophysical context (including extreme magnetic fields of magnetars; \citet{beskin16}) allow to employ the test-field approximation which neglects its effect on the geometry of the spacetime and Maxwell equations are solved independently to determine the electromagnetic field. 

Curved spacetimes of compact objects (black holes or neutron stars) may substantially deform the electromagnetic field in its neighborhood and several purely relativistic effects arise. In particular, in the case of extremal rotating black hole any external axisymmetric magnetic field is expelled  from the event horizon. Expulsion of the field lines is known as black hole Meissner effect and it was originally discussed for particular test-field solutions and later also for several exact solutions describing magnetized black holes \citep{bicak00,karas91,bicak85,wald74}, and recently it has been further generalized using the formalism of weakly isolated horizons \citep{gurlebeck18,gurlebeck17}.

While the axisymmetry is crucial for the Meissner effect to operate, other types of relativistic effects may appear if we consider non-axisymmetric systems of magnetized compact objects.  In particular, it has been shown that rotating Kerr black hole set in uniform motion in external asymptotically  homogeneous magnetic field misaligned with the spin axis creates extremely complicated structure of field lines leading to the close contact of the lines of anti-parallel orientation and even to the formation of X-type null points  \citep{karas09}. Magnetic null points are typically associated with the process of magnetic reconnection occurring in plasma and presence of charged matter and electric currents is essential for their emergence in classical magnetohydrodynamics. However, it appears that relevant structure of magnetic field may be formed due to relativistic effects of frame-dragging and spacetime curvature even in the electro-vacuum magnetospheres \citep{karas14,karas13,karas12}. 

More recently, the vacuum magnetosphere of a neutron star in the vicinity of a supermassive black hole was considered in this context. In particular, it has been shown that magnetic null points may form even in the Rindler approximation of this system \citep{kopacek18b}. Rindler limit neglects the spacetime curvature, which is justified in the very vicinity of the black hole horizon, and gravitation of the static black hole is represented solely by the acceleration \citep{macdonald85}. Rindler approximation is consistent with the final stages of the plunging trajectory until the neutron star reaches the horizon of the central massive black hole. While the formation of the magnetic nulls within the magnetosphere could support the release of energy leading to the acceleration of charged matter and high-power electromagnetic emission, the scenario in which a stellar mass compact object is inspiralling and finally plunges into supermassive black hole (i.e., extreme mass ratio inspiral; EMRI) also represents a promising source of gravitational waves for the future space-based observatories like LISA \citep{babak17}.

In the previous paper \citep{kopacek18b}, we employed  the Rindler approximation to find the solution of Maxwell equations for the plunging neutron star idealized as a rotating conducting spherical source of dipolar magnetic field arbitrarily inclined with respect to the axis of rotation. We discussed the solution in near zone (without the radiative terms) and, in particular, we found that magnetic null points may emerge within such magnetosphere. Nevertheless, the system was treated in geometrized units scaled by the mass of the central black hole and the consistency with the parameters of realistic astrophysical systems has not been verified. In this contribution we discuss physical values of parameters and check whether the formation of magnetic null points in the magnetosphere is indeed astrophysically relevant.

\section{Plunging neutron star} \label{NS}
We consider a neutron star at the final stage of its inspiral close to the horizon of the supermassive Schwarzschild black hole. Near-horizon region of the Schwarzschild spacetime may be approximated by the flat Rindler spacetime \citep{orazio13,macdonald85,rindler66} with metric given in Rindler coordinates  $(t,x,y,z)$ as follows:
\begin{equation}
ds^2 = -\alpha^2 dt^2 + dx^2 + dy^2 +dz^2,
\label{rindler_metric}
\end{equation}
where the lapse function $\alpha$ is given as $\alpha=g_{H}z$ and $g_{H}$ denotes the horizon surface gravity. We use dimensionless geometrized units where the speed of light $c=1$.

\begin{figure}[htb]
\begin{center}
\includegraphics[width=.5\linewidth]{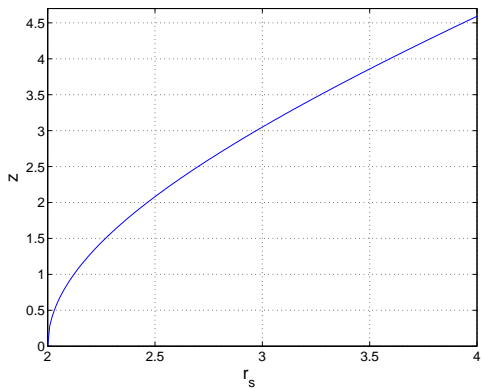}
\end{center}
\caption{\label{z_vs_r}Rindler coordinate $z$, measuring the proper distance from the event horizon of the black hole, as a function of the Schwarzschild radial coordinate $r_s$.}
\end{figure}

Relation between Rindler coordinates and  Minkowski coordinates $(T,X,Y,Z)$ is given by the transformation:

\begin{equation}\label{Rindler transf}
T=z\,\sinh(g_{H}t),\,\,\,X=x,\,\,Y=y,\,\,\,Z=z\,\cosh(g_{H}t).
\end{equation} 

Rindler coordinate $z$ measures the proper distance from the horizon:
\begin{equation}\label{z_r}
z=\int_{2}^{r_s}\frac{\dif{r}}{\sqrt{1-2/r}}=\log{\left(\frac{\sqrt{1-2/r_s}+1}{|\sqrt{1-2/r_s}-1|}\right)}+r_s\,\sqrt{1-2/r_s},
\end{equation}
where $r_s$ is the Schwarzschild radial coordinate scaled by the rest mass of the black hole $M$ (i.e., $M=1$ is set in all equations). Rindler horizon at $z=0$ corresponds to the Schwarzschild event horizon at $r_s=2$ and Rindler approximation of the Schwarzschild spacetime thus remains appropriate for sufficiently small $z$ (e.g., to keep $r_s \lessapprox 2.5$ demands $z \lessapprox 2$). The relation (\ref{z_r}) between $z$ and $r_s$ is plotted in Fig.~\ref{z_vs_r}.

\begin{figure}[ht]
\begin{center}
\includegraphics[width=.3\linewidth]{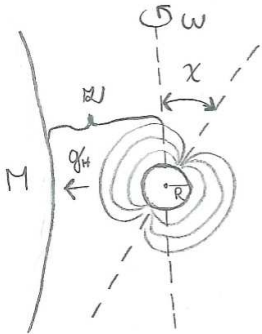}
\end{center}
\caption{\label{sketch}Sketch of the investigated model (not to scale). Neutron star of the radius $R$ is rotating with the angular frequency $\omega$ and its dipole-type magnetic field  is inclined by angle $\chi$ with respect to the rotation axis. The neutron star is plunging into the nearby horizon of the supermassive black hole with mass $M$ and the proper distance from the horizon is given by the Rindler coordinate $z$. In the adopted near-horizon approximation, the gravitational effects of the black hole are fully characterized by the acceleration $g_H$.}
\end{figure}

We consider a vacuum magnetosphere of a superconducting neutron star of radius $R$ and rotation frequency $\omega$ as a source of dipolar magnetic field with the inclination angle $\chi$ with respect to the rotation axis. The neutron star is free-falling from its initial position at $(0,0,Z_s)$ towards the horizon. The plunge is parametrized by the Rindler coordinate time $t$ and the star's position in Rindler coordinates evolves as $(0,0,Z_s/\cosh{g_H t})$. Sketch of the model is presented in Fig.~\ref{sketch}. Spatial distance from the dipole is expressed in Rindler coordinates as follows:
\begin{equation}
r=\sqrt{x^2+y^2+(z\,\cosh(g_{H}t)-Z{s})^2},
\end{equation} 
and retarded time $\tau$ is given as:
\begin{equation}\label{proper time}
\tau=T-r=z\,\sinh(g_{H}t)-\sqrt{x^2+y^2+(z\,\cosh(g_{H}t)-Z{s})^2}\,.
\end{equation}

Resulting electromagnetic field for the freely falling rotating magnetic dipole was derived in \citet{kopacek18b} while the similar setup was previously considered by \citet{orazio13}. In the near zone (dropping all radiative terms) we obtain following components of the magnetic field vector expressed in Rindler coordinates for the observer co-moving with the Rindler frame:

{\footnotesize
\begin{eqnarray}
	B_{x}&=&\frac{m}{r^5}\left\{\cosh(g_{H}t)\Bigg[\sin(\chi)\left\{3\,x\left[x\,\cos(\omega\tau)+y\,\sin(\omega\tau)\right]-r^2\,\cos(\omega\tau)\right\}+3\,(z\,\cosh(g_{H}t)-Z_{s})\,x\,\cos(\chi)\Bigg] \right.\nonumber \\
	&&-\omega\sinh(g_{H}t)\left[(z\,\cosh(g_{H}t)-Z_{s})\sin(\chi)\left\{\frac{5\,R^2\,y}{r^2}\left(x\,\cos(\omega\tau)+y\,\sin(\omega\tau)\right)+(r^2-R^2)\,\sin(\omega\tau)\right\} \right. \nonumber \\
	&&\left. \left.+R^2\,y\,\cos(\chi)\left(\frac{5\,(z\,\cosh(g_{H}t)-Z_{s})^2}{r^2}-1\right)\right] \right\}, \label{Magn-x-Cond-Rind}\\
	B_{y}&=&\frac{m}{r^5}\left\{\cosh(g_{H}t)\Bigg[\sin(\chi)\left\{3\,y\left[x\,\cos(\omega\tau)+y\,\sin(\omega\tau)\right]-r^2\,\sin(\omega\tau)\right\}+3\,(z\,\cosh(g_{H}t)-Z_{s})\,y\,\cos(\chi)\Bigg] \right.\nonumber\\
	&&+\omega \sinh(g_{H}t) \left[(z\,\cosh(g_{H}t)-Z_{s})\sin(\chi)\left\{\frac{5\,x\,R^2}{r^2}\left(x\,\cos(\omega\tau)+y\,\sin(\omega\tau)\right)+(r^2-R^2)\,\cos(\omega\tau)\right\}\right. \nonumber \\
	&&\left. \left.+R^2\,x\,\cos(\chi)\left(\frac{5\,(z\,\cosh(g_{H}t)-Z_{s})^2}{r^2}-1\right)\right]\right\}, \label{Magn-y-Cond-Rind} \\
	B_{z}&=&\frac{m}{r^5}\Bigg[3\,(z\,\cosh(g_{H}t)-Z_{s})\,\sin(\chi)\,\left[x\,\cos(\omega\tau)+y\,\sin(\omega\tau)\right]+\cos(\chi)\left(3\,(z\,\cosh(g_{H}t)-Z_{s})^2-r^2\right)\Bigg], \label{Magn-z-Cond-Rind}
\end{eqnarray}
}%
where $m$ is the magnitude of the dipole moment.

\section{Magnetic null points}\label{NP}
Magnetic null points (NPs) are locations within the magnetosphere where the components of the magnetic field (\ref{Magn-x-Cond-Rind})-(\ref{Magn-z-Cond-Rind}) simultaneously vanish, i.e., $(B_x,B_y,B_z)=(0,0,0)$. We have numerically confirmed that NPs may develop in the employed model of the magnetosphere and identified following  necessary conditions for their existence in given setup: (i) non-zero acceleration ($g_H>0$); (ii) inclination of the dipole $\chi\neq 0, \pi/2$; and (iii) rotation of the dipole $\omega>0$  \citep{kopacek18b}. Moreover, we discussed how the emergence and the position of the NP depends on the Rindler time $t$ and radius of the neutron star $R$ for several values of inclination $\chi$. Regarding the former, we were able to numerically locate the NP only for some period of coordinate time $t$ which slightly differed for each $\chi$. For fixed $t$ we investigated the effect of radius $R$. We found that presence of conducting sphere is not crucial for the formation of NPs, which were located also for $R=0$. With increasing value of $R$, the location of NP changes and may approach the surface of the star, however, it always remains outside ($r>R$).

\begin{figure}[htb]
\begin{center}
\includegraphics[width=.495\linewidth]{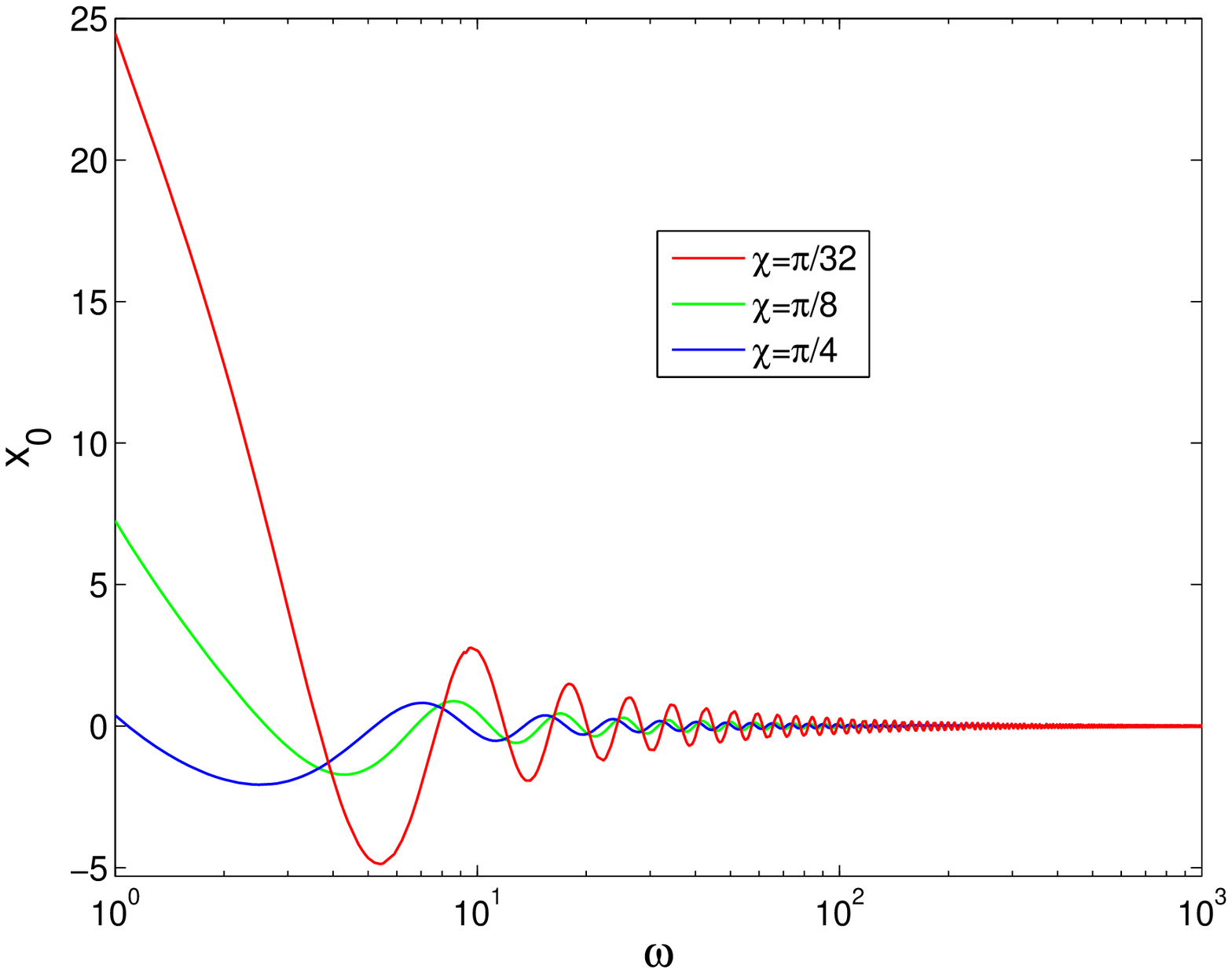}
\includegraphics[width=.495\linewidth]{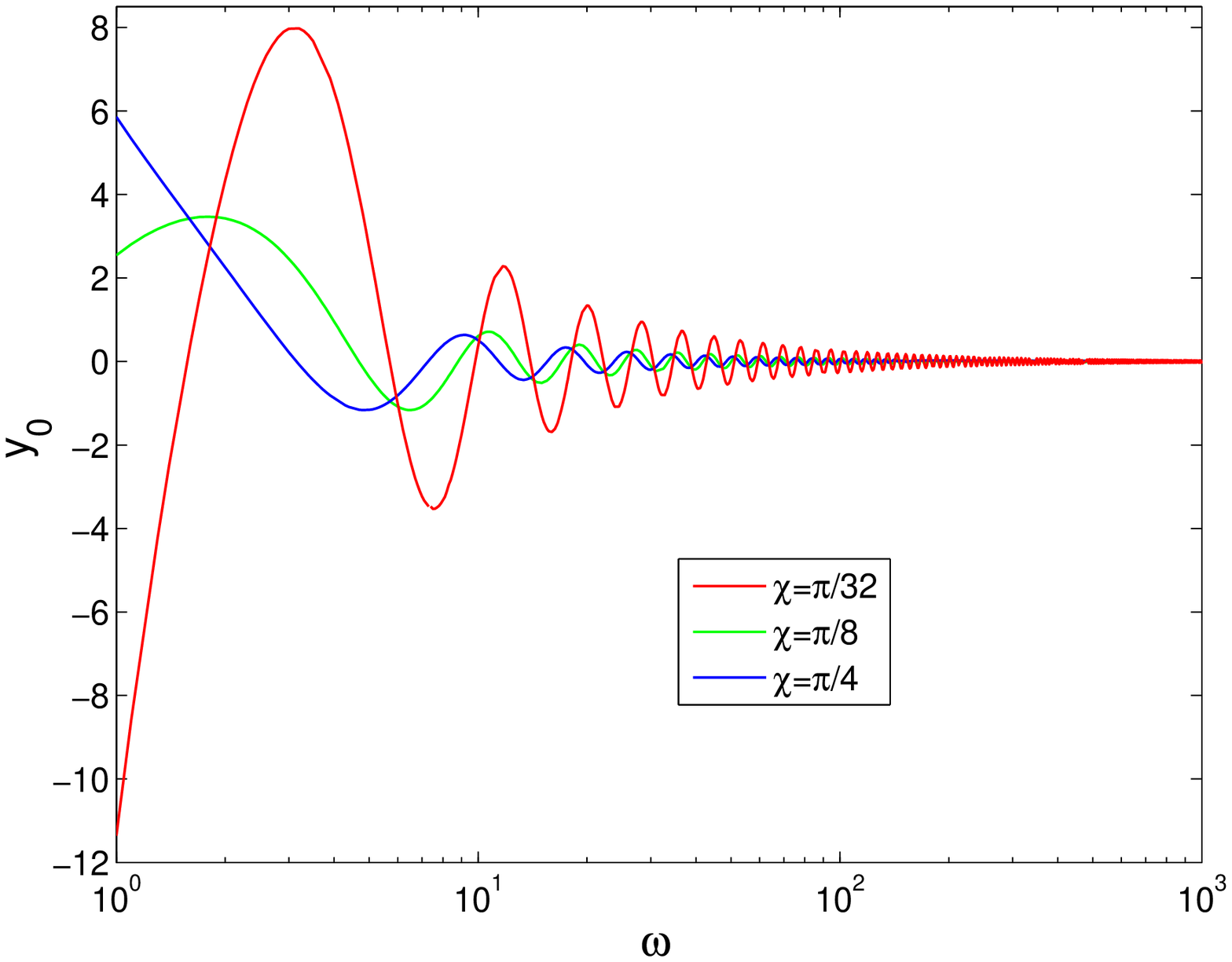}\\
\includegraphics[width=.495\linewidth]{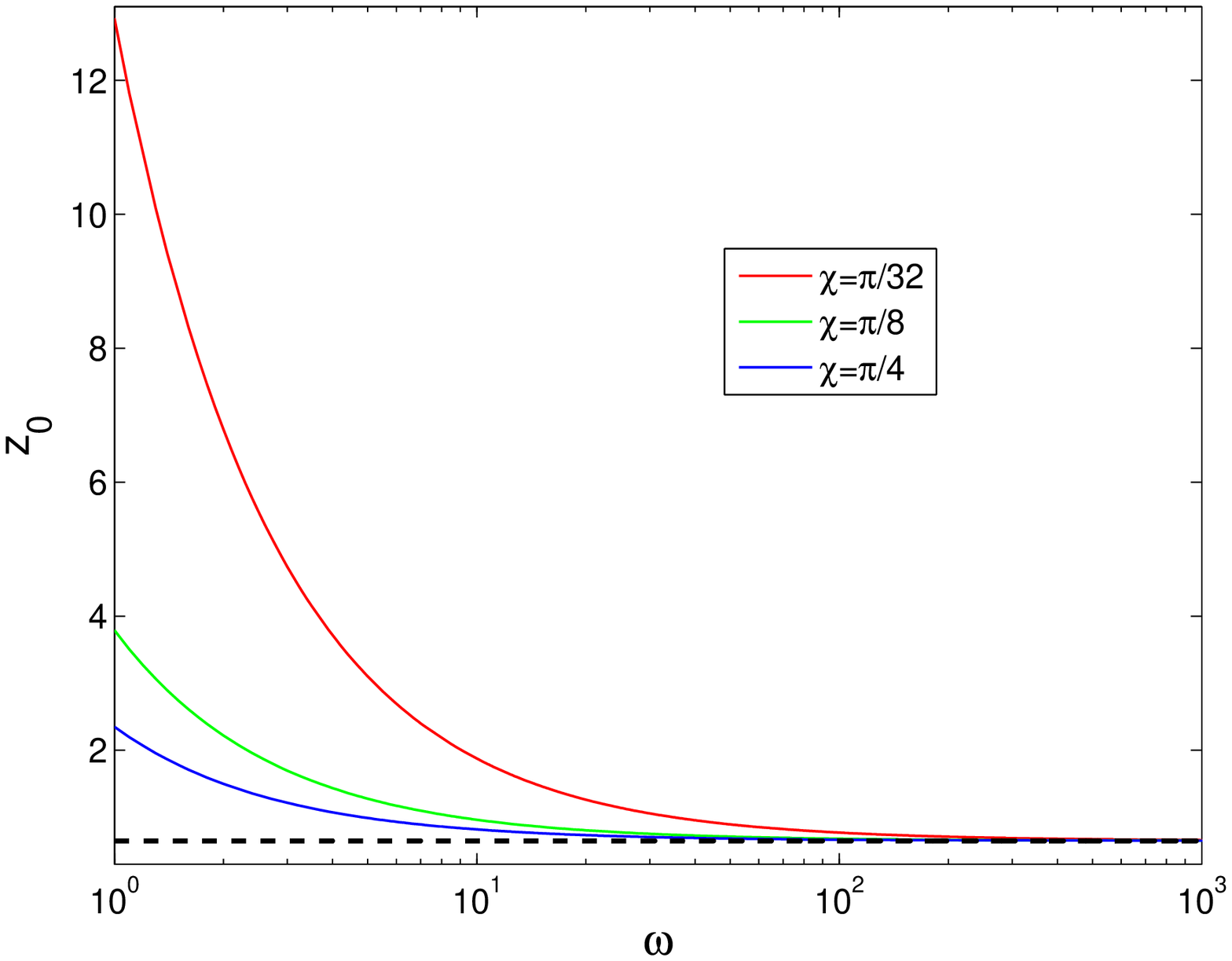}
\includegraphics[width=.495\linewidth]{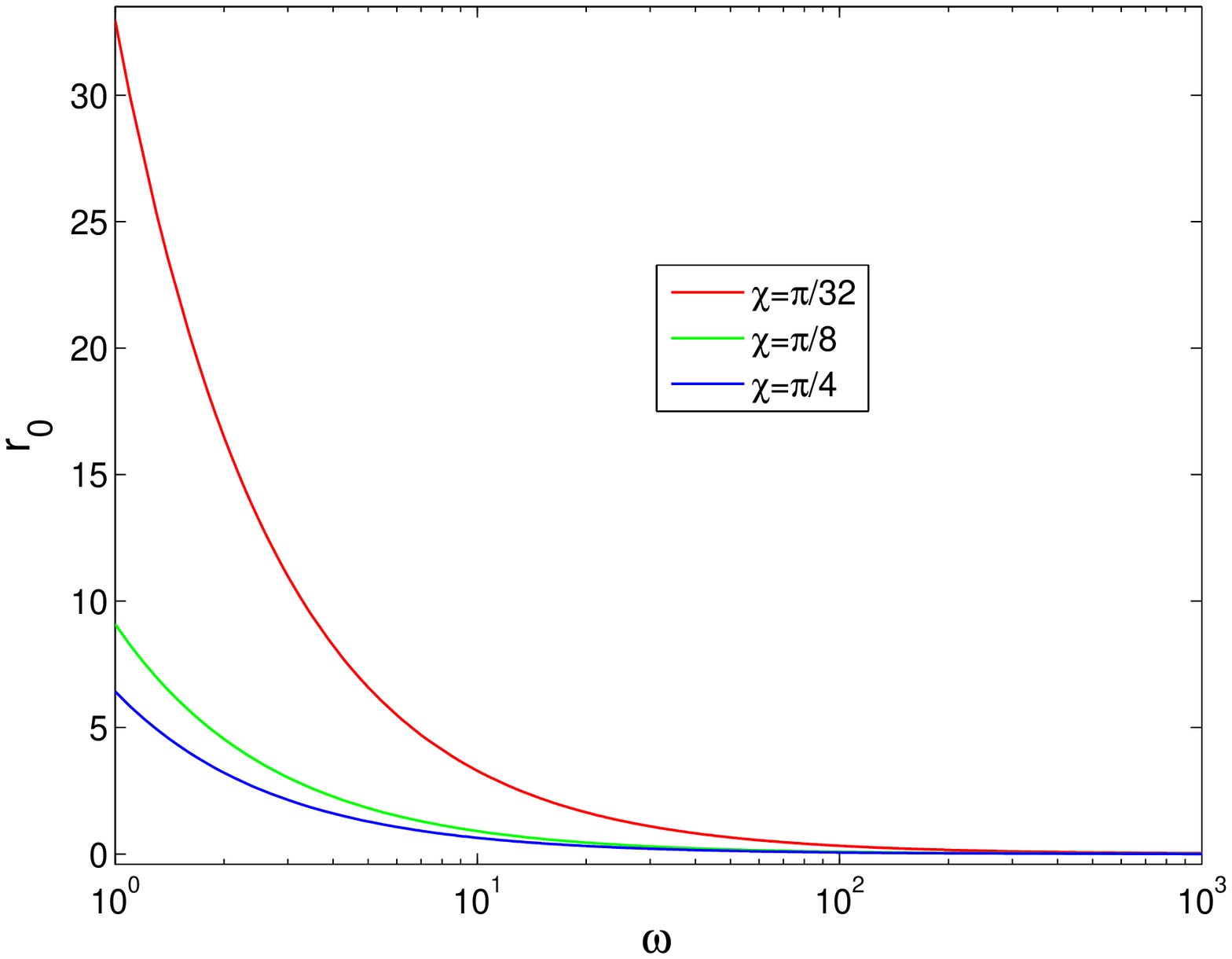}
\end{center}
\caption{\label{diskuze_omega}Rindler coordinates $x_0$, $y_0$, $z_0$ and the distance from the dipole $r_0$ of the magnetic null point as a function of the rotation frequency $\omega$ for several values of the inclination angle $\chi$. Dashed line in the bottom left panel indicates the current location of the plunging neutron star at $z=0.648$. Remaining parameters  of the model are fixed as: $Z_s =1$, $g_H t=1$ and $R=10^{-5}$.}
\end{figure}

In the previous analysis we discussed the formation of NPs and their locations with respect to the parameters in geometrized dimensionless units scaled by the rest mass of the central black hole $M$. In this contribution we intend to verify the consistency of observed effects with realistic astrophysical system of neutron star plunging into supermassive black hole.

Rotation of the neutron stars is detected directly in pulsars with observed periods in the range $P_{\rm SI}\approx 10^{-3}-10 \;\rm{s}$ \citep{hessels06,tan18} and angular frequency in SI units $\omega_{\rm{SI}}=2\pi/P_{\rm SI}$ is related to its dimensionless value $\omega$ as follows:  
\begin{equation}\label{freq}
\omega=\frac{ \omega_{\rm{SI}}\;(1472\;\rm{m})}{c}\left(\frac{M}{M_{\odot}}\right),
\end{equation}
where the factor $1472\;\rm{m}$ is the value of solar mass $M_\odot$ in geometrized units.

Radius of the neutron star is $R_{\rm SI}\approx 10\,\rm{km}$ and its value $R$ in dimensionless units is given as :
\begin{equation}\label{radius}
R=\frac{R_{\rm{SI}}}{(1472\;\rm{m})}\left(\frac{M_\odot}{M}\right).
\end{equation}

For the central black hole we consider a mass range of $M\approx 10^6 - 10^9 M_\odot$. The lower mass limit yields the dimensionless frequency in the range $\omega \approx 3-3\times 10^4$ while the upper limit leads to $\omega \approx 3\times 10^3 - 3\times 10^7$.  In the previous analysis \citep{kopacek18b} we fixed the frequency as $\omega=1$ which is, however, below the relevant astrophysical range, and discussed the role of remaining parameters. Here we complete the discussion and study the effect of increasing $\omega$ on the formation and location of the NP in the magnetosphere.

\begin{figure}[htb]
\begin{center}
\includegraphics[width=.85\linewidth]{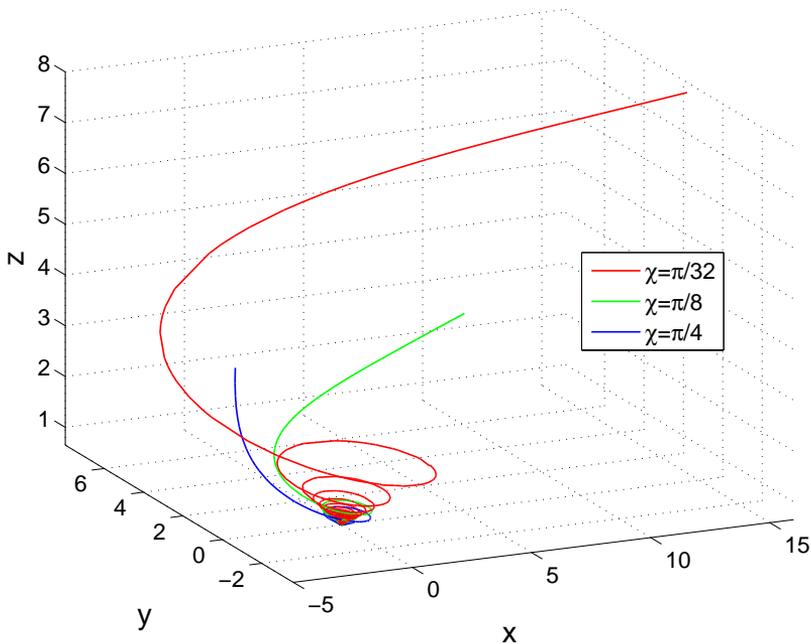}
\end{center}
\caption{\label{diskuze_omega_3d}Locations of the magnetic null points with varying frequency $\omega$ for several values of the inclination angle $\chi$. Magnetic null points gradually approach the neutron star located at $(x,y,z)=(0,0,0.648)$ as $\omega$ increases. Same data sets as in Fig.~\ref{diskuze_omega} are presented here.}
\end{figure}

Iterative root-finding routine is applied to numerically locate NPs of the field (\ref{Magn-x-Cond-Rind})-(\ref{Magn-z-Cond-Rind}) with sufficient precision. In Fig.~\ref{diskuze_omega} we present Rindler coordinates $x_0$, $y_0$, $z_0$ and the distance $r_0$ of the NP as a function of $\omega$ for several values of inclination $\chi$. It shows that the NP gradually approaches the neutron star as $\omega$ increases and that for each $\omega$ the NP is always closer for higher inclinations. Locations of the same set of NPs are presented in 3D plot in Fig.~\ref{diskuze_omega_3d} which shows the gradual {\em inspiral} of the NP to the vicinity of the neutron star as the rotation frequency rises.  

The values of remaining parameters  of the model were fixed as: $Z_s =1$, $g_H t=1$ and $R=10^{-5}$. Given value of $R$ corresponds to the mass $M\approx10^6\;M_\odot$ set in Eq.~(\ref{radius}). In agreement with previous results we observe that NPs approach the neutron star but always remain above its surface. The value of the initial location of the star $Z_s=1$ corresponds to the Schwarzschild radial coordinate $r_s \approx 2.1$ which is consistent with the Rindler approximation. The choice of $g_H t=1$ does not put any astrophysical constraint as the Rindler coordinate time $t$ is a free parameter which parametrizes the plunge.

\begin{figure}[htb]
\begin{center}
\includegraphics[width=.85\linewidth]{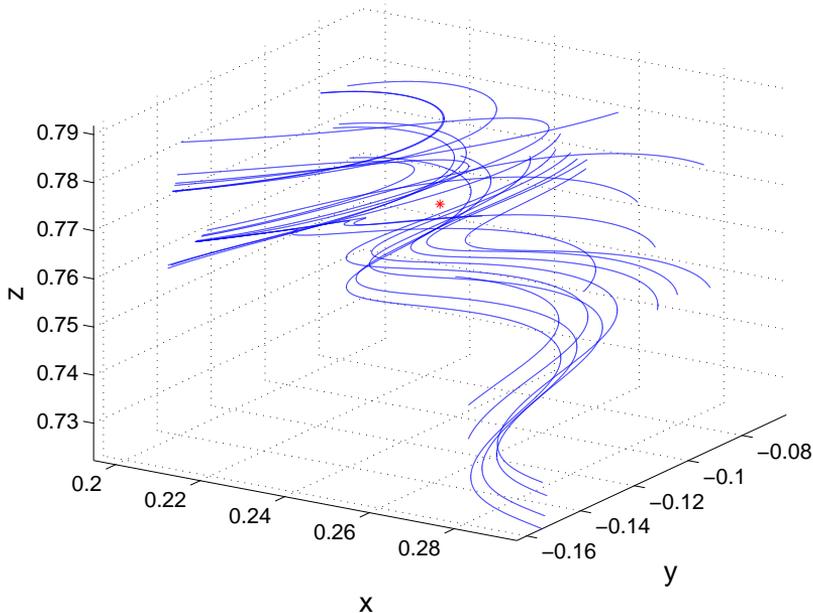}
\end{center}
\caption{\label{NP_3D}Magnetic field lines in the vicinity of the null point (red mark) located at $x_0=0.245$, $y_0=-0.114$ and $z_0=0.771$. The following values of parameters are set: $Z_s =1$, $g_H t=1$, $\omega=100$, $\chi=\pi/32$ and $R=10^{-5}$.}
\end{figure}

The behavior of the magnetic field close to the NP with $\omega=100$ and $\chi=\pi/32$ is shown in Fig.~\ref{NP_3D}. Structure of the field lines in this region becomes very complicated and the field intensity changes rapidly on the small spatial scale. With higher $\omega$ the variability of the field in the close vicinity of the neutron star further increases and makes the structure of the field lines too complex for the visual inspection. 

For the same reason, the numerical method used to locate NPs encounters increasing difficulties for $\omega\gtrsim 1000$. However, the behavior for $\omega \leq 1000$ observed in Figs.~\ref{diskuze_omega} and \ref{diskuze_omega_3d} suggests that NPs would further approach the surface of the neutron star. With high spin frequencies, $\omega > 1000$, we expect to find the NPs in the immediate neighborhood of the neutron star\footnote{The formation of the NP within the superconducting interior of the star is not possible as demonstrated in previous paper \citep{kopacek18b}.}, which is located near the horizon of the central black hole and Rindler approximation may thus be applied to describe this region of the magnetosphere.

\section{Summary}
Locations of magnetic null points which emerge in the electro-vacuum magnetosphere of the neutron star near the supermassive black hole were discussed. We verified the astrophysical relevance of the investigated scenario and completed our previous analysis. In particular, we studied the role of spin frequency $\omega$ and found that realistic values of $\omega$ generally allow the formation of the NP close to the neutron star, which guarantees the consistency with employed Rindler approximation of the Schwarzschild spacetime. 

The results suggest that during the final stages of the inspiral, the strong gravity effects of central black hole support the release of electromagnetic energy in the process of magnetic reconnection leading to the acceleration of charged particles and powerful emission of electromagnetic radiation from the magnetosphere of the infalling neutron star.

\ack
This work was supported from the following grants of the Grant agency of the Czech republic: No. 17-06962Y (O. K.), No. 17-13525S (T. T.) and No. 19-01137J (V. K.). We acknowledge the Inter-Excellence COST project LTC 18058 of the Czech Ministry of Education, Youth and Sports.

\bibliography{\jobname}
\end{document}